# High-Temperature Photocurrent Mechanism of *β*-Ga$_2$O$_3$ Based MSM Solar-Blind Photodetectors

B. R. Tak,[1] Manjari Garg,[1] Sheetal Dewan,[2] Carlos G. Torres-Castanedo,[3] Kuang-Hui Li,[3] Vinay Gupta,[2] Xiaohang Li[3] and R. Singh[1]

[1]*Department of Physics, Indian Institute of Technology Delhi, New Delhi-110016, India*

[2]*Department of Physics and Astrophysics, University of Delhi, Delhi-110007, India*

[3]*King Abdullah University of Science and Technology (KAUST), Advanced Semiconductor Laboratory, Thuwal, 23955-6900, Saudi Arabia*

High-temperature operation of metal-semiconductor-metal (MSM) UV photodetectors fabricated on pulsed laser deposited *β*-Ga$_2$O$_3$ thin films has been investigated. These photodetectors were operated up to 250 °C temperature under 255 nm illumination. The photo current to dark current (PDCR) ratio of about 7100 was observed at room temperature (RT) while it had a value 2.3 at 250 °C at 10 V applied bias. A decline in photocurrent was observed from RT to 150 °C and then it increased with temperature up to 250 °C. The suppression of the blue band was also observed from 150 °C temperature which indicated that self-trapped holes in Ga$_2$O$_3$ became unstable. Temperature-dependent rise and decay times of carriers were analyzed to understand the photocurrent mechanism and persistence photocurrent at high temperatures. Coupled electron-phonon interaction with holes was found to influence the photoresponse in the devices. The obtained results are encouraging and significant for high-temperature applications of *β*-Ga$_2$O$_3$ MSM deep UV photodetectors.

**Keywords:** High temperature responsivity, Solar blind photodetector, Ga$_2$O$_3$, persistent photocurrent, activation energy

Deep UV (DUV) photodetectors also known as solar-blind photodetectors with high thermal stability have drawn considerable attention due to wide potential applications of defense, security, space communications, UV astronomy, biological and environmental studies.[1-5] As the science advances, high-temperature DUV photodetectors will be an important constituent for the aforementioned applications.[5-6] For high-temperature applications, a photodetector must fulfill the requirements of high thermal stability, high chemical stability, high-temperature operation, high electrical tolerance and high spectral selectivity. Most of the UV photodetectors in commercial applications are based on silicon (Si) due to well established Si-technology. However, UV photodetectors based on current Si-based technology possess limitations of high dark current due to narrow bandgap and require wood's filters.[7] Additionally, heavy cooling systems were also required for traditional Si-photodetectors to achieve promising external quantum efficiency. For high-temperature applications, these photodetectors are unable to function due to large thermally generated carriers which led to negligible photoresponse. In the past few years, wide bandgap semiconductor materials such as AlGaN, SiC, and ZnMgO based solar-blind photodetectors were explored for operation at elevated temperatures.[8-14] The wide bandgap of these materials was advantageous and eliminate the requirement of wood's filters. Heavy doping in AlGaN and ZnMgO causes crystalline quality degradation and suffer from background signals in the environment.[15-16] The AlGaN/GaN photodetectors were reported at elevated temperatures of 200°C.[8] However, the photo to dark current ratio (PDCR) (~ 0.1) and decay time (34 s) at 200 °C was not so impressive. A faster decay time was reported at 200 °C than at RT due to faster electron-hole recombination process. SiC-based photodetectors possess limitations such as cost and indirect band gap which motivated researchers for the search of new materials.[1]

Nowadays, an upsurge in research on $\beta$-Ga$_2$O$_3$ based DUV photodetectors has been stimulated due to its intrinsic solar blind nature and high absorption coefficient (>10$^5$).[17-24] The intrinsic solar blind nature eliminated the complexity of doping which is advantageous over AlGaN, SiC, ZnMgO etc. It also possesses eminent thermal and chemical stability which makes it a potential contender to function in harsh environments.[25-26] However, there are only few reports available on high-temperature operation of solar blind photodetectors.[9, 27-28] Recently, S. Ahn et al. reported the temperature dependent PDCR of Si-implanted $\beta$-$Ga_2O_3$ thin films grown by metal organic chemical vapor deposition (MOCVD) with Ti/Au metal contacts.[27, 29] They observed a continuous increase in photocurrent with temperature due to the presence of defects in the bandgap. These photodetectors showed high persistence photocurrent even at RT due to the ohmic behavior of Ti/Au contacts on $Ga_2O_3$. T. Wei et al. also demonstrated the potential of $Ga_2O_3$ photodetectors at 427 °C using IZO as a transparent electrode.[30] However, they performed measurements under 185nm illumination with a very high power density of 14 W/m$^2$. Also, in the previous reports,[8, 27-28, 30-32] photoresponse on various high-temperature stable UV photodetectors was shown to either increase or decrease with increase in temperature. The physical mechanism responsible for photoresponse increase or decrease with temperature was not explained. Therefore, the development of solar blind photodetectors for high-temperature applications are at their early stage and needs further investigations to understand the physical mechanism of photocurrent transport and persistence photocurrent at high temperatures.

Herein, the high-temperature performance of DUV photodetectors fabricated on $Ga_2O_3$ thin films was investigated at different temperatures from RT up to 250 °C. The $Ga_2O_3$ thin films used for fabrication of photodetectors were deposited using pulsed laser deposition (PLD) technique. X-ray diffraction (XRD) and transmission electron microscopy (TEM) with fast Fourier transform (FFT) were performed to analyze the crystalline quality of as-deposited thin films. The low dark current is desired to obtain high PDCR. Therefore, the metal-semiconductor-metal (MSM) structure was fabricated with Ni/Au as a Schottky metal contact. Temperature-dependent spectral responsivity, PDCR and temporal response characteristics exhibited superior performance of $Ga_2O_3$ solar-blind photodetectors for operation in harsh environment. The charge carrier transport mechanism across the metal-semiconductor interface is also discussed with respects to spectral and temporal responses which deepen the understanding of the operation of solar-blind photodetectors at high temperatures. Combined with the cheaper deposition process, obtained results of $Ga_2O_3$ solar-blind photodetectors are encouraging and paves the way for the aforementioned high-temperature applications.

The $Ga_2O_3$ thin films used for fabrication of DUV photodetectors were deposited on sapphire (0001) substrates using PLD. The deposition was carried out at 800 °C substrate temperature. The 200 mJ KrF excimer laser with 10 Hz frequency was used for the ablation of gallium oxide target. The oxygen pressure was maintained at $5\times10^{-4}$ Torr during the deposition. Further experimental information can be found in reference.[33] X-ray diffraction (Philips Xpert Pro) having Cu Kα (λ = 1.54 Å) radiation was used for structural investigation of as-grown $Ga_2O_3$ thin film. Transmission Electron Microscopy (TEM) images were recorded using Titan $G^2$ 80-300 ST system with a line resolution of 0.1 nm.

Further, interdigitated electrodes on $Ga_2O_3$ thin film were patterned in Class-100 clean room using maskless lithography system (Intelligent micropatterning, SF-100). These electrodes were 700 µm long and 50 µm wide with figure spacing of 50 µm. The metal contacts of Ni (30 nm)/ Au (40 nm) were deposited using a thermal evaporation system. Figure S1 (supplementary file) illustrates the schematic of the experimental setup used for studying the temperature dependent properties of on *β*-$Ga_2O_3$ DUV photodetectors. A nitrogen purged Xenon lamp (75W) combined with a computer interfaced monochromator (Bentham TMC-300V) was used for the spectral responsivity measurements. An optical UV Fiber (PCU-1000) was used to direct monochromatic beam over the device under test (DUT) placed on a high-temperature DC probe station (Ever-Being International Corporation, EB 6). The temperature was controlled using a temperature controller ranging from RT to 300°C. The power spectrum of the Xenon lamp was acquired using a Thorlabs power meter (PM-100D) and a calibrated Si-photodiode (S-130VC). A Keithley semiconductor parameter analyzer (SCS-4200) was also connected for external biasing.

Figure 1(a) depicts the typical X-ray diffraction (XRD) 2θ-scan of gallium oxide thin film. The peaks marked with a star (*) corresponds to the c-plane sapphire substrate. The XRD results revealed that the monoclinic structure of $\beta$-Ga$_2$O$_3$ thin film was epitaxially grown in (-201) orientation. Cross-sectional transmission electron microscopy (CS-TEM) and the Fast Fourier Transform (FFT) measurements were performed to further investigate the crystalline quality and Ga$_2$O$_3$-Al$_2$O$_3$ interface as shown in Figure 1(b) and 1(c) respectively. The sharp Ga$_2$O$_3$-Al$_2$O$_3$ interface (< 2nm) as depicted in Figure 1(b) indicated a very small lattice mismatch which makes sapphire a preferred substrate for the growth of Ga$_2$O$_3$ thin films. The FFT image indicated good crystalline quality of Ga$_2$O$_3$ thin films.

The scanning electron microscope (SEM) image of the fabricated device is shown in the inset of Figure 1(d) on a 100 µm scale. The interdigitate MSM structure was fabricated with two back to back Ni/Au Schottky contacts. The MSM photodetectors exhibit low dark current and high speed than photodiodes.[34] The RT dark current of 3×10$^{-10}$ A was obtained even at 10V bias as shown in Figure S2(a) (Supplementary file). The low dark current was achieved due to back to back Ni/Au Schottky contacts. It was 3-4 times lower than the reported dark current for Ohmic metal contacts on Ga$_2$O$_3$.[29, 35] The low dark current even at high applied voltage bias is desirable for superior photodetector performances. Figure S2(b) (Supplementary file) depicts the variation in photocurrent with temperature. It was observed that the photocurrent was varied lesser than the dark current with temperature increment. Photocurrent to dark current (PDCR) ratio is also an essential parameter of photodetectors which was calculated using the following equation:[36-37]

$$PDCR = \frac{I_p - I_d}{I_d} \qquad (1)$$

where $I_p$ and $I_d$ are denoted as photocurrent and dark current respectively. The temperature dependent PDCR is shown in Figure 2(a). Photocurrent measurements were performed at 10 V applied bias under 255 nm illumination with a power density of 400 µW/cm². It was observed that Ga$_2$O$_3$ MSM photodetectors can be used to detect DUV light up to a high temperature of 250°C due to high thermal stability. The PDCR value was diminished from 7137 to 2.3 as the temperature was increased from 25°C to 250°C. The intrinsic dark current level was again obtained after cooling the device from high temperature to RT which showed that Ga$_2$O$_3$ based Ni/Au MSM DUV photodetectors possess excellent thermal stability. The variation in PDCR values at the highest operating temperature was observed to be better as compared to the previous reports on AlGaN, SiCN and Ga$_2$O$_3$ based UV detectors which are depicted in Table 1. The maximum operating temperature was also higher than the AlGaN and SiCN UV detectors. Although the Si-doped Ga$_2$O$_3$ photodetector showed operation at higher temperatures, but it suffered high persistence photocurrent even at room temperatures.[29, 38] In our case, the high-temperature operation only up to 250°C was performed due to system limitations. The decrement in PDCR values as temperature raised was resulted due to the increased dark current which was resulted from thermally generated charge carriers.

To analyze the thermal stability at high temperatures, variable temperature photocurrent and responsivity measurements of fabricated DUV photodetector with 255nm UV light illumination were analyzed as shown in Figure 2(b). Initially, both photocurrent and peak responsivity were decreased up to 150°C, but as the temperature was increased further, both the values started to increase. The responsivity of photodetectors is defined as [39-40]

$$R_\lambda = \frac{I_p - I_d}{P.S} \quad (2)$$

where P is the power density of illuminated light and S is the effective device area which is 0.008 cm² for DUT. The peak responsivity of 0.70 A/W was obtained at RT and it decreased to 0.40 A/W at 150 °C. It increased to 0.74 A/W as the temperature further increased to 250°C. The decrease in photoresponsivity with temperature was also observed in earlier reports.[28, 31-32, 41] Enhancement of responsivity is desirable for the performance of DUV detectors at high temperatures. The normalized spectral responsivity with varying temperature from RT to 250°C is depicted in Figure 3(a). It can also be observed from temperature dependent spectral response that the blue band suppressed as temperature increased. The origin of blue band in $Ga_2O_3$ was attributed to recombination of electrons to self-trapped holes (STH).[42-45] The quenching of blue band was reported above 400K. In this report, the suppression of the blue band was observed from 150 to 250°C. Hence, variable temperature spectral response indicated that the STH were no longer stable and become mobile at high temperatures.

UV/visible discrimination ratio is another important performance parameter of photodetectors. The RT UV (255 nm)/visible (500 nm) discrimination ratio was about 4 orders of magnitude and it decreased to about 2 orders of magnitude at 250°C. Large UV/visible discrimination ratio at high temperatures assures the practical application of fabricated DUV photodetector.

The temporal response of MSM photodetector was investigated to evaluate the detection speed. Time-dependent photocurrent measurements were performed under illumination of 255 nm wavelength at 10 V applied bias. The rise and decay time of photocurrent were qualitatively analyzed by fitting the temporal response curve with following bi-exponential equation:[46-47]

$$I = I_0 + A_1 e^{-t/\tau_1} + A_2 e^{-t/\tau_2} \qquad (3)$$

where $I_0$ is the steady-state current, $\tau_1$ and $\tau_2$ are the relaxation time constants and $A_1$ and $A_2$ are constants. In general, rise and decay times consist of two components having a fast and a slow response. The fast response component was assigned to change in carrier concentration upon illumination switching. However, the slow response component was attributed to intrinsic defects influenced charge carrier trapping/de-trapping.[39] Real-time current change under switching ON/OFF of UV light illumination with varying temperature from RT to 250°C was recorded as shown in Figure S3 (Supplementary file). All of these current pulses were fitted with Equation (3) and corresponding rise and decay time constants were obtained as shown in Figure 3(b). It was observed that the fast component of decay time ($\tau_{d1}$) remained constant below 150°C and then it increased with the rise in temperature. However, fast rise time component ($\tau_{r1}$) initially decreased from 590 to 163 ms as the temperature was varied from 25 to 150°C. It was then slightly increased to 193 ms when the temperature was raised to 250 °C. Evolution of slow components showed that rise time $\tau_{r2}$ decreased from 2.4 to 1.4 s with an increase in temperature from 25 to 150 °C then it increased to 3.2 s when the temperature reached 250 °C. However, slow decay time component $\tau_{d2}$ continuously increased from 1.6 to 5.4 s with continuous variation of temperature up to 250 °C.

To understand the physical mechanism of photocurrent behavior with temperature, the slow components of rise time ($\tau_{r2}$) was plotted with 1/T as depicted in figure 4(a).[48-49] The activation energies of 73 and 205 meV were obtained from the linear fit of the Arrhenius plot of rise time $\tau_{r2}$ below and above 150°C respectively. The physical meaning of negative activation energy is related to the carrier generation/recombination ratio.[50] This ratio is less than one for negative activation energy which showed that carrier recombination mechanism is the dominant process. The positive activation energy was attributed to the effective carrier generation mechanism. The low activation energy of 73 meV was corresponding to the energy of electron-phonon interactions in $Ga_2O_3$.[51-52] The coupled electron-phonon with low energy of 73 meV undergoes scattering phenomenon.[52] Therefore, phonon-assisted trapping of electrons with STH occurred below 150 °C due to electron scattering which resulted in the formation of self-trapped excitons (STE). Hence, trapping of charge carriers was responsible for the decrement in the photocurrent. The activation energy of 205 meV was corresponding to the transition of STH to the mobile hole in valence band which resulted in fewer electrons trapping.[53] Hence, the photocurrent was increased for temperatures above 150 °C due to mobile holes. This conclusion was also supported by variable temperature spectral response measurements as discussed earlier.

To understand the origin of PPC in Ga$_2$O$_3$ photodetectors, the slow component of decay time ($\tau_{d2}$) was plotted with temperature. The Arrhenius plot of $\tau_{d2}$ is shown in figure 4(b). The two activation energies of 58 meV and 168 meV were obtained below and above 150 °C respectively. These energy barriers have been proposed to the energy of electron-phonon interaction and transition of mobile holes to STH in the gallium oxide respectively.[51, 53] Hence, in the lower temperature regime, the electron-hole recombination occurs through thermally generated phonons. The mobile holes require additional energy to capture electrons above 150 °C. Hence, the high PPC was attributed to less electron interaction with hole due to higher energy of coupled electron-phonon and mobile holes at high temperatures. Schematic in figure 4(c) and (d) dictates the photocurrent mechanism below and above 150 °C.

In conclusion, high-temperature stability of MSM photodetectors on $β$-Ga$_2$O$_3$ epitaxial films up to 250°C was demonstrated. The UV/Visible discrimination ratio was as high as $10^4$ at RT and $10^2$ at 250°C which offered high-temperature performance of Ga$_2$O$_3$ solar-blind photodetectors. The photoresponsivity initially decreased from RT to 150°C and then increased with further enhancement in temperature up to 250°C. The PPC was also increased in the temperature range from RT to 250°C. The obtained activation energies from Arrhenius plot of the rise and decay times were corresponding to the energy of electron-phonon interaction and transition of STH to mobile holes in $β$-Ga$_2$O$_3$. Hence, carrier generation and recombination mechanisms were found to occur via electron-phonon interaction with holes. The excellent properties of Ga$_2$O$_3$ photodetectors pave the way for next-generation high temperature stable solar-blind photodetectors.

**ACKNOWLEDGMENTS**

The authors (BRT and RS) would like to thank Department of Physics, IIT Delhi for providing XRD facility. We would also like to acknowledge Nanoscale Research Facility, IIT Delhi for device fabrication and characterizations. Department of Science and Technology (DST), India is highly appreciated for awarding INSPIRE research fellowship to BRT for the PhD programme. The KAUST authors are thankful for the supports of KAUST baseline BAS/1/1664-01-01, KAUST CRG URF/1/3437-01-01 and GCC Research Council REP/1/3189-01-01.

# Figures

**Figure 1 (a) XRD 2θ scan of pulsed laser deposited β-Ga$_2$O$_3$ thin film (b) Cross-sectional TEM (CS-TEM) of Ga$_2$O$_3$/Al$_2$O$_3$ interface (c) Fast Fourier Transform (FFT) pattern of β-Ga$_2$O$_3$ thin film(d) SEM image of fabricated photodetector at 100 μm scale.**

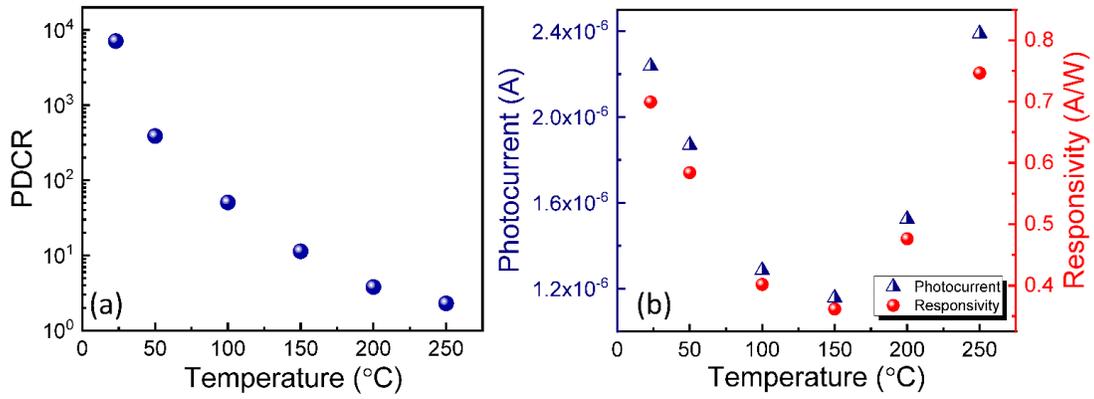

**Figure 2 (a) Temperature-dependent PDCR (b) photocurrent and peak responsivity of β-Ga$_2$O$_3$ photodetector at 10V bias. and 255 nm illumination**

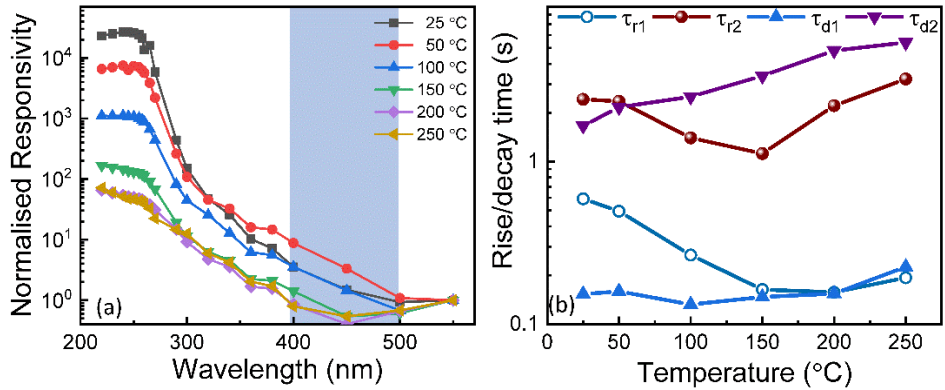

**Figure 3 (a) Spectral response of fabricated photodetector with temperature variation ranging from 23°C to 250°C (b) Rise times ($\tau_{r1}$ and $\tau_{r2}$) and decay times ($\tau_{d1}$ and $\tau_{d2}$) under 255 nm illumination with detector temperature**

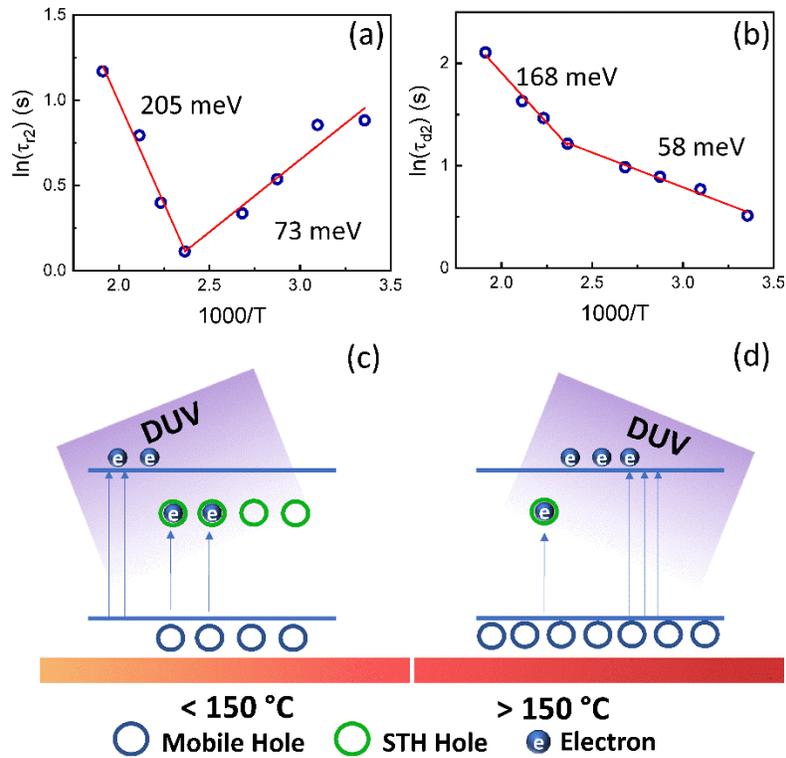

**Figure 4** Arrhenius plot of slow components of (a) rise and (b) decay time (c) & (d) shows schematic representation of photocurrent generation and recombination mechanisms below and above 150 °C

**Tables**

**Table 1** High-temperature performance of solar blind photodetectors

| Photodetector | Wavelength (cm) | Maximum operating Temperature $T_{max}$ (°C) | PDCR at $T_{max}$ | UV power density (μW/cm$^2$) | Reference |
|---|---|---|---|---|---|
| AlGaN | 275 | 150 | - | - | 28 |
| SiCN | 254 | 200 | 2.3 | 500 | 9 |
| Si-doped β-Ga$_2$O$_3$ (MOCVD) | 254 | 350 | 9 | 760 | 27 |
| β-Ga$_2$O$_3$ (MOCVD) | 185 | 427 | 1.5 | 14000 | 30 |
| β-Ga$_2$O$_3$ (PLD) | 255 | 250 | 2.3 | 400 | This work |